\documentclass[fleqn,usenatbib]{mnras}

\usepackage{newtxtext,newtxmath}
\usepackage[T1]{fontenc}
\usepackage{ae,aecompl}

\usepackage{graphicx}	% Including figure files
\usepackage{amsmath}	% Advanced maths commands
\usepackage{amssymb}	% Extra maths symbols
\usepackage{amssymb}
\usepackage{natbib}
\usepackage[parfill]{parskip}
\usepackage{enumitem}
\usepackage[mathscr]{euscript}

\numberwithin{equation}{section}

\title[Bondi-Parker Flow]{Stability and Solution of the Time-Dependent Bondi-Parker Flow}

\author[Keto]{
Eric Keto$^{1}$\thanks{E-mail: eketo@cfa.harvard.edu (EK)}
\\
$^{1}$Institute for Theory and Computation, Harvard College Observatory, 60 Garden St., Cambridge, MA 02138\\
}

\date{Accepted XXX. Received YYY; in original form ZZZZ}

\pubyear{2019}

\begin{document}
\label{firstpage}
\pagerange{\pageref{firstpage}--\pageref{lastpage}}
\maketitle

\begin{abstract}
\citet{Bondi1952} and \citet{Parker1958} derived a steady-state solution for Bernouilli's equation in spherical symmetry around a point mass for two cases, respectively, an inward accretion flow and an outward wind. Left unanswered were the stability of the steady-state solution, the solution itself of time-dependent flows, whether the time-dependent flows would evolve to the steady-state, and under what conditions a transonic flow would develop. In a Hamiltonian description, we find that the steady state solution is equivalent to the Lagrangian implying that time-dependent flows evolve to the steady state. We find that the second variation is definite in sign for isothermal and adiabatic flows, implying at least linear stability. We solve the partial differential equation for the time-dependent flow as an initial-value problem and find that a transonic flow develops under a wide range of realistic initial conditions. We present some examples of time-dependent solutions.
\vskip 0.25truein
\end{abstract}


%%%%%%%%%%%%%%%%%%%%%%%%%%%%%%%%%%%%%%%%%%%%%%%%%%

%%%%%%%%%%%%%%%%% BODY OF PAPER %%%%%%%%%%%%%%%%%%
\section{Introduction}

Spherical accretion onto a constant point mass \citep{Bondi1952} has found application in astronomy
from stars to supermassive black holes and
anywhere the self-gravity of the gas is insignificant compared to the gravity of the point mass. 
The simplicity of the model combined with the non-trivial solution
that includes a transonic critical point, 
has proven both useful and interesting.  The same
equations but with outward velocities \citep{Parker1958} has found application in solar and stellar winds and
anywhere acceleration occurs as a result of a pressure-density
gradient maintained by a gravitational field. The acceleration in the Parker wind is also the same as occurs 
through a rocket nozzle with an exponential shape. 

The Bondi-Parker (BP) flow is described by a combination of the continuity equation with Bernouilli's equation,
the latter a partial 
differential equation (PDE) for velocity and density as functions of time and position. Assuming the time derivative is zero
results in a single ordinary differential equation (ODE) for the steady state with separable variables. The
transcendental equation resulting from integration is easily solved, for example with a Newton-Raphson technique
or in terms of the Lambert W function (Cranmer 2004). The constant of integration along with the branches of
a quadratic term results in
a family of steady-state trajectories with either subsonic, transonic, or supersonic velocities.

There have been many interesting variations of the BP flow.
For example, the introduction of shock discontinuities
links different trajectories in the steady-state family \citep{McCrea1956}. A non-isothermal equation of
state results in multiple critical points \citep{Kopp1976}. Accounting for the self-gravity of the gas results in a similarity
solution either as
a function of time alone \citep{Shu1977} or time and radius \citep{Dhang2016} depending on the equation of state and the
initial conditions.

The stability of the steady-state solution has previously been
addressed with finite-difference methods \citep{Balazs1972,Stellingwerf1978,Garlick1979,Velli1994,DelZanna1998}. 
These studies agree
that the transonic flow is stable, but disagree about the stability of the
subsonic and supersonic flows based on 
differences in numerical methods and choice
of boundary conditions. 

In contrast, a Hamiltonian description of the flow determines the evolution and stability
independently of numerical methods and boundary conditions. We find that the steady-state solution is
equivalent to the Lagrangian for the flow and is thus the critical function for the first variation of the functional
of the flow. This condition implies that time-dependent flows evolve to the steady-state/
We find that the second variation is definite in sign for isothermal and adiabatic equations of state.
This implies that these time-dependent flows are at least linearly stable. 

The choice of trajectory in the steady-state family that results from the evolution is determined by
the initial velocities. We use the method of characteristics to write the PDE for the time-dependent
BP flow as a pair of coupled ODEs 
for velocity and position both as functions of time. These may be  solved numerically as an initial
value problem (IVP), for
example with a Runge-Kutta technique. These solutions specify the final trajectory for any initial conditions.

Examples suggest that the time-dependent solutions evolve to a transonic flow
from all initial values that lie within the region bounded above and below by the inward and outward steady-state
transonic trajectories.
Since this region extends asymptotically from velocities with absolute values
from zero to infinity, a wide range of initial velocities results in a transonic flow.

\section{The steady state Bondi-Parker flow}\label{SSBP}
\begin{figure}
\includegraphics[width=3.25in]{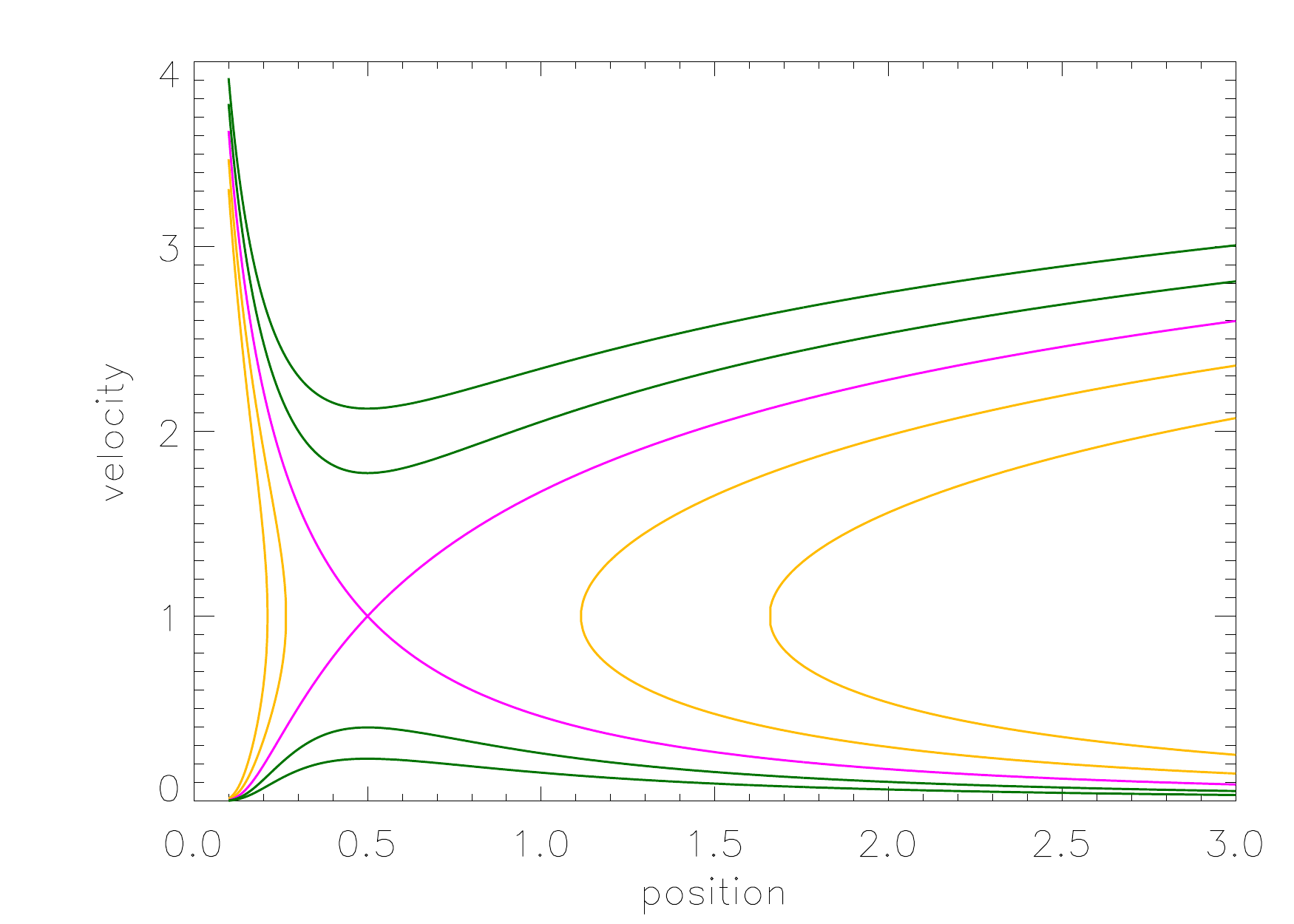}
\caption{
Representative trajectories of the steady-state BP flow (equation \ref{ss}) in non-dimensional units. 
The velocities are shown as absolute values.
The two transonic trajectories with $\mathscr{L}=\mathscr{L}_C$ are shown in pink. Green lines show two subsonic trajectories with
$\mathscr{L} = \mathscr{L}_C -2$ and $\mathscr{L}=\mathscr{L}_C-1$ and two supersonic trajectories with 
$\mathscr{L} = \mathscr{L}_C +1$ and $\mathscr{L}=\mathscr{L}_C+2$. 
The trajectories that are discontinuous across position are shown in yellow for the same set of constants, $\mathscr{L}$
but represent different branches.
}
\label{bondiparker_plot}
\end{figure}

Following \citet{Bondi1952} and \citet{Parker1958} we derive the steady-state solution.
If the gas is isothermal so that  $\partial P / \partial \rho = a^2$ for sound speed $a$, and
the gravitational force is that of a point of constant mass, $M$, then the Euler equation in spherical symmetry is,
\begin{equation}\label{tdbpeq}
\frac {\partial \tilde{u}} {\partial \tilde{t}} = -\tilde{u} \frac{\partial \tilde{u}}{\partial \tilde{x}} - \frac{a^2}{\tilde{\rho}} \frac{\partial \tilde{\rho}}{\partial \tilde{x}} - \frac{GM}{\tilde{x}^2}
\end{equation}
where the tilde indicates a variable with dimensional units.
This can be written in non-dimensional form with the definitions,
\begin{equation}\label{scaling}
\tilde{x} = \bigg(\frac{GM}{a^2}\bigg) x, \ \ \tilde{u} = au,\ \ \tilde{\rho} = \tilde{\rho_1}\rho  ,\ \ {\rm and} \ \ \tilde{t} = \bigg(\frac{GM}{a^3}\bigg) t,
\end{equation}
where $\tilde{\rho_1}$ is an arbitrary density.
With these substitutions, equation \ref{tdbpeq} is,
\begin{equation}\label{ndbpeq}
\frac {\partial u}{\partial t} = -u \frac {\partial u} {\partial x} - \frac {1} {\rho}   \frac {\partial \rho}{\partial x} - \frac{1}{x^2}.
\end{equation}
The density may be eliminated with the help of the non-dimensional continuity equation,
\begin{equation}\label{continuity}
\rho = \lambda x^{-2}u^{-1}
\end{equation}
where $\lambda$ is the accretion rate. Then,
\begin{equation}\label{tdndeq}
\frac{\partial u}{\partial t} = \bigg( \frac{1}{u} - u\bigg) \frac {\partial u}{\partial x} + \bigg(\frac{2}{x} - \frac{1}{x^2}\bigg).
\end{equation}
In steady state the time derivative on the left-hand side is zero, and
the variables can be separated
and integrated,
\begin{equation}\label{ss}
 \mathscr{L} = \log |u| - \frac{1}{2} u^2  + 2\log |x| + \frac{1}{x} .
\end{equation}
In non-dimensional units, the constant of integration $\mathscr{L}=\log \lambda$ is equivalent to the energy and
with a simple non-linear scaling to the 
the mass accretion rate.
This transcendental equation can be solved either numerically, for example with a Newton-Raphson technique, 
or using the Lambert W function (Cranmer 2004), 
\begin{equation}
u = \pm \bigg\{-W\bigg[ -\exp \bigg(\frac{2}{x} + 4 \ln |x| + 2\mathscr{L}\bigg) \bigg]\bigg\}^{1/2}
\end{equation}
or
\begin{equation}
x = -2W \bigg[ \pm \frac{1}{2} \exp \bigg(\frac{1}{2} u^2 -\ln| u| + \mathscr{L}\bigg) \bigg].
\end{equation}
The family of solutions depends on the value of the integration constant, $\mathscr{L}$, and upon the branches of
the quadratic term or the Lambert W function. 
Representative solutions
are plotted in figure \ref{bondiparker_plot}. 
Of particular interest are the two transonic solutions, the Bondi accretion flow and the Parker wind, 
both with $\mathscr{L} = \mathscr{L}_C = 3/2 - 2\log 2$ that 
cross at the Bondi-Parker critical point, $(x,u) = (\frac{1}{2},1)$ or $(\tilde{x},\tilde{u}) = (GM/(2a^2), a)$. 
Also shown are
subsonic flows with $\mathscr{L} < \mathscr{L}_C$, 
and supersonic flows with $\mathscr{L}>\mathscr{L}_C$.  The
solutions that are discontinuous in position derive from a different branch and
are not
accretion flows or winds. The region of the plot that they occupy is sometimes called the forbidden region.
A more complete discussion is found in \citet{Cranmer2004} and \citet{Holzer1970}.

% FIGURE 1

\section{Hamiltonian Description}

Hamilton's principle for a conservative system is,
\begin{equation}
\delta J = \delta \int^{t_2}_{t_1} (K-V) dt = 0
\end{equation}
where the difference of the kinetic and potential energies is the Lagrangian, $K-V = \mathscr{L}$.
This may be generalized for a fluid with internal energy, $U$ \citep{Herivel1955},
\begin{equation}\label{lagrangian}
\mathscr{L} = K-V-U.
\end{equation}
In non-dimensional units, the Lagrangian for the BP flow is the same as the steady-state equation, \ref{ss},
where the non-dimensional terms, $(\log |u| + 2\log |x|)$, that derive from the
dimensional pressure term in equation \ref{tdbpeq}, are the internal energy. 
If we define the functional gradient as the directional derivative,
\begin{equation}
\nabla J[x]  = \frac{d}{d\epsilon} J[x + \epsilon\eta] \bigg\vert_{\epsilon=0},
\end{equation}
for any variation, $\eta$, and constant, $\epsilon$,
then the necessary condition of Euler-Lagrange for $x$ to be a critical function of $J[x]$ is that the first variation is zero
\begin{equation}
\langle \nabla J[x] \rangle = \frac {\partial}{\partial x} \mathscr{L}(x,\dot{x},t) - \frac{d}{dt} \bigg( \frac {\partial}{\partial \dot{x}} \mathscr{L}(x,\dot{x},t) \bigg)
\end{equation}
where the velocity $u = \dot{x}$. 

Since the right side is equivalent to the equation of motion \ref{tdndeq} as can be verified by
substitution, then $\langle \nabla J[x] \rangle= 0$,
and the steady-state
solution, $\mathscr{L}$, is the solution of stationary action toward which the time-dependent solutions evolve.

This description is similar to the simple physics problem in which the Lagrangian 
determines the parabolic trajectory of a particle in a
gravitational field and the exact trajectory is determined  from the end points as a boundary value problem (BVP) 
or alternatively from the initial
velocities as an IVP. However, the Lagrangian for the BP flow includes the internal energy of the fluid as a function of 
the density-pressure gradient which is constrained by the conservation equation. 
Therefore, it is not always possible to find an energy conserving trajectory between
two arbitrary velocities. Considered as a BVP, not all boundary conditions are consistent with a steady-state solution.

As in Lagrangian mechanics, the energy serves as a Lyapunov function for the stability.
The definiteness in sign of the second variation evaluated at the steady
state  implies at least  
linear stability  \citep{Arnold1965,Arnold1989}.
Equivalently, the integrand of the second variation,
\begin{equation}
\eta^2 \frac{\partial^2 \mathscr{L}}{\partial x^2}  + 2\eta\dot{\eta} \frac{\partial^2 \mathscr{L}}{\partial x \partial \dot{x}}+ \dot{\eta}^2 \frac{\partial^2 \mathscr{L}}{\partial \dot{x}^2}
\end{equation}
must be definite in sign for every nonzero variation $\eta$. Since the middle term with mixed partial derivatives is zero, and
\begin{equation}\label{L1}
 \frac{\partial^2 \mathscr{L}} {\partial x^2} = -2\bigg( \frac{1}{x^2} + \frac{1}{x^3} \bigg)
\end{equation}
\begin{equation}\label{L2}
\frac{\partial^2 \mathscr{L} }{\partial \dot{x}^2} = - \bigg( 1 + \frac{1}{\dot{x}^2} \bigg).
\end{equation}
this is easily verified for the
range $x > 0$
allowed for BP flows.
This implies that the steady-state solution, including all the subsonic, transonic, and supersonic trajectories in the family, 
is at least linearly stable. 

The existence of a steady-state solution, a critical point in Lagrangian mechanics, 
is required for stability. An arbitrary IVP can evolve to the steady-state, but the
boundary conditions in a BVP must be consistent with the conservation of energy to
allow a stable solution. This requirement itself does not determine the steady-state solution or
its stability. For example, the derivation of the Euler-Lagrange equation assumes that the functional to be
maximized or minimized has values equal to two specified endpoints.  In between, the solution
depends on the functional. 

\section{The time dependent Bondi-Parker flow}\label{TDBP}
To solve for the time-dependent BP flow, equation \ref{tdndeq} we use
the method of characteristics to write the PDE as a pair of coupled first-order 
ODE's
that may be solved as an initial value problem.  Equation  \ref{tdndeq} written as,
\begin{equation}\label{conventional}
\frac{d[u(t,x(t))]}{dt} = \frac{\partial u}{\partial t} + \bigg(u - \frac{1}{u} \bigg)  \frac{\partial u}{\partial x} = \frac{2}{x} - \frac{1}{x^2},
\end{equation}
and compared with the identity,
\begin{equation}\label{identity}
\frac{du}{dt} = \frac{\partial u}{\partial t} \frac{dt}{dt} + \frac{\partial u} {\partial x} \frac{dx}{dt},
\end{equation}
suggests the pair of coupled ODE's,
\begin{equation}\label{ode1}
\frac{dx}{dt} = u - \frac{1}{u}
\end{equation}
\begin{equation}\label{ode2}
\frac{du}{dt} = \frac{2}{x} - \frac{1}{x^2}
\end{equation}
to be solved with initial values, $u(t,x) = f(x)$ at $t=0$.
Alternatively, these may be derived from the Lagrange-Charpit (LC) equations for the PDE \ref{tdndeq},
\begin{equation}\label{lc}
\frac{dt}{1} = \frac{dx}{u-u^{-1}} =\frac{du}{2x^{-1} - x^{-2}}.
\end{equation}
The LC equations also yield a third ODE that is equivalent to the steady-state equation \ref{ss}.

Following the method of characteristics, we parameterize the ODE's with 
functions $t=g(\tau,s)$ and $x=h(\tau,s)$ to obtain the following set of ODE's with their initial conditions
\begin{equation}\label{lc1}
\frac{dt}{d\tau} = 1 \ \ {\rm with} \ \ t(0,s) = 0,
\end{equation}
\begin{equation}\label{lc2}
\frac{dx}{d\tau} = u - \frac{1}{u} \ \ {\rm with} \ \ x(0,s) = s,
\end{equation}
\begin{equation}\label{lc3}
\frac{du}{d\tau} = \frac {2} {x} - \frac{1}{x^2}   \ \ {\rm with} \ \  u(0,s) = f(s).
\end{equation}
Here $(\tau,s)$ are the initial values for the trajectories, $x(t),u(t)$.
This set of coupled ODE's may be solved numerically, for example with a Runge-Kutta technique for $u(\tau,s)$
To complete the solution for $u(t,x)$, we need the inverse functions,
$t=g(\tau,s)$ and $x=h(\tau,s)$. 
From equation \ref{lc1}, $\tau$
is identical to $t$. From a set of solutions of equations \ref{lc2} and \ref{lc3} we determine $x=h(t,s)$ 
to obtain the solution $u(t,h(t,s)) = u(t,x)$.

\section{Examples}\label{EXAMPLES}

Example solutions suggest the proposition that a transonic flow develops if any of the
initial velocities are within the forbidden region. Because the
forbidden region extends asymptotically over $0 < |u| < \infty$, there is a wide range of initial values 
that will result in a transonic flow. Any constant nonzero initial velocities will cross a boundary of the
forbidden region if the range in position is large enough. Stated another way,  an initial flow has to be 
constructed in a special way to avoid the forbidden regions. 
The velocities of the subsonic and supersonic steady-state trajectories have this property.

While generalizing from examples is short of a mathematical proof, this proposition seems plausible on physical grounds.
Consider an inflow with a subsonic velocity in the forbidden region with $x  < x_C$. This flow will be accelerated to 
approximate free fall,  $u\sim 1/ \sqrt{x}$, by the gravitational force. Equation \ref{tdndeq} constrains the velocity of the steady-state solution 
to the sound speed at the critical point, and the flow will evolve until this condition is met.
The critical point thus acts as an outer boundary condition for the supersonic region of the inflow. The
critical point is also an inner
boundary condition for the subsonic region of the flow, $x > x_C$. Here the flow is able to adjust to the 
transonic solution by increasing the density and 
pressure to slow the flow until its transonic point occurs at the critical point, and
both terms on the right-hand side of equation \ref{tdndeq} are zero.
This description also applies
to the wind with appropriate modifications.

The transonic critical point effectively provides a natural boundary condition for the 
inward and outward transonic flows.
In the case of a stellar wind for example, if an arbitrary radius is chosen for the
base of the wind, then the velocity at the base is uniquely determined by the 
requirement that the wind pass through the transonic critical point.
In the case of an astrophysical subsonic or supersonic flow, while 
any single point in the $(x,u)$ plane 
(figure \ref{bondiparker_plot}) uniquely determines one steady-state trajectory,
this would have to be determined outside of the BP flow, for example
by supposing that conditions in the interstellar medium set a constant
velocity at a particular radius. This velocity would need to be outside
the forbidden region, assuming the proposition above.

\subsection{Transonic solutions}\label{CONSTANT}

\begin{figure*}
\begin{tabular} {p{3in}c}
\includegraphics[width=3in]{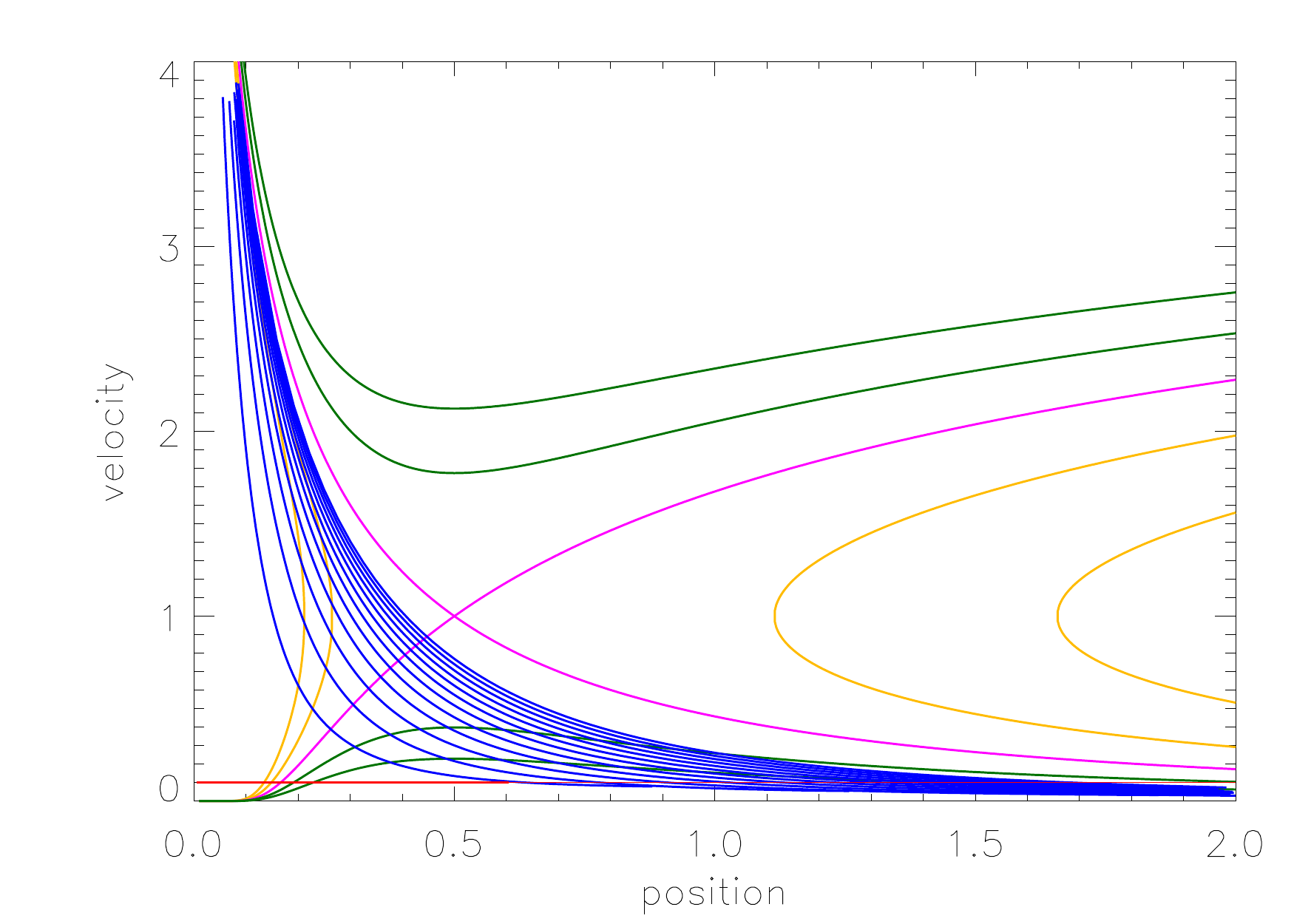}
&
\includegraphics[width=3in]{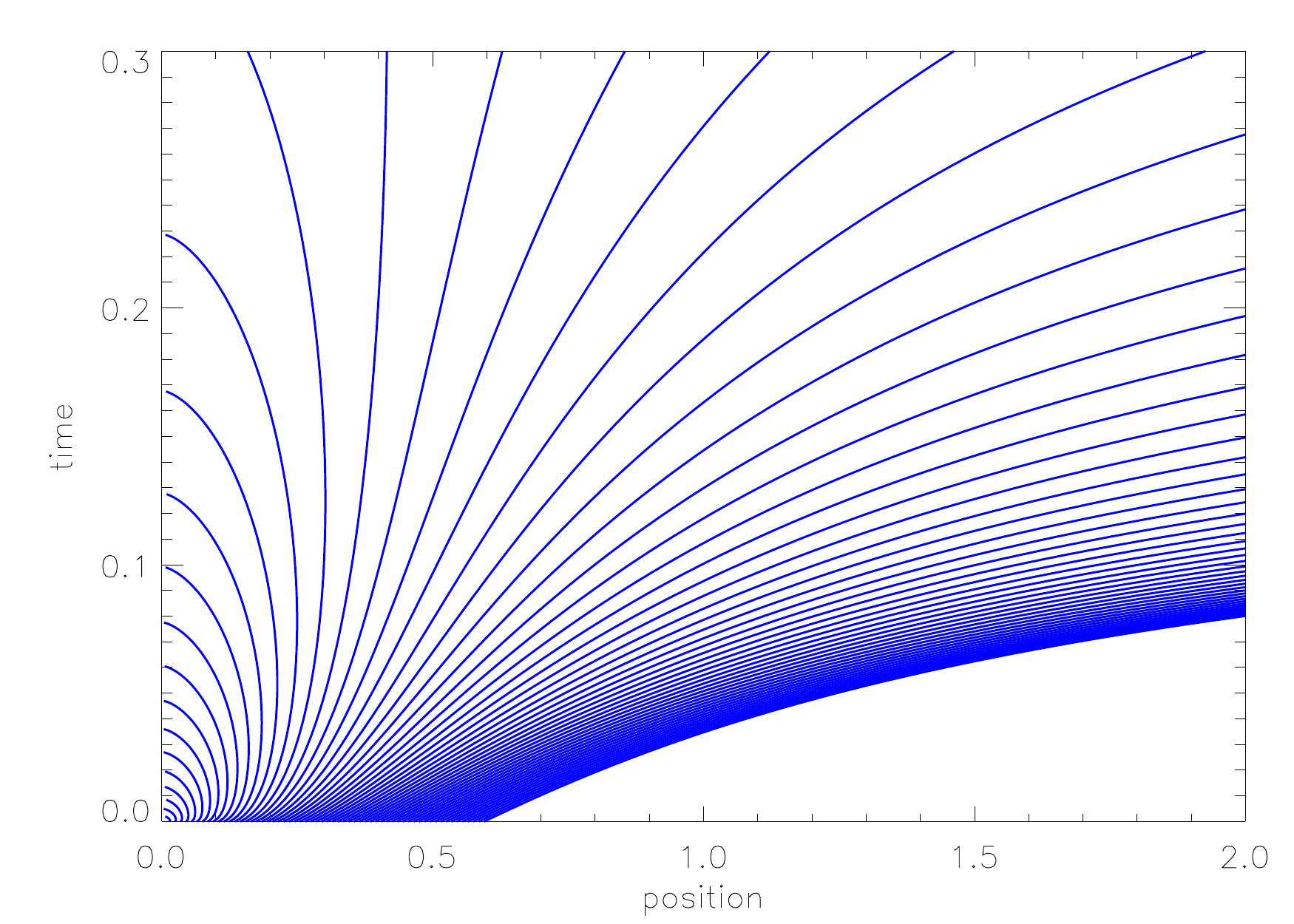}
\end{tabular}
\caption{
Left: Time evolution of an accretion flow with initial values $u(x) = -0.2$ and $0.005 <x < 2.0$ plotted on top of
the steady state trajectories from figure \ref{bondiparker_plot}. The time-dependent solution is shown in blue 
for a sequence of times with the initial values in red.
Velocities are plotted as their absolute values. The flow evolves asymptotically to the steady-state transonic trajectory,
the Bondi accretion flow.
Right: Characteristics for the solution. Only the characteristics originating in the range (0.005 < x < 0.6) are shown.
}
\label{bondisub_plot}
\end{figure*}

\begin{figure*}
\begin{tabular} {p{3in}c}
\includegraphics[width=3in]{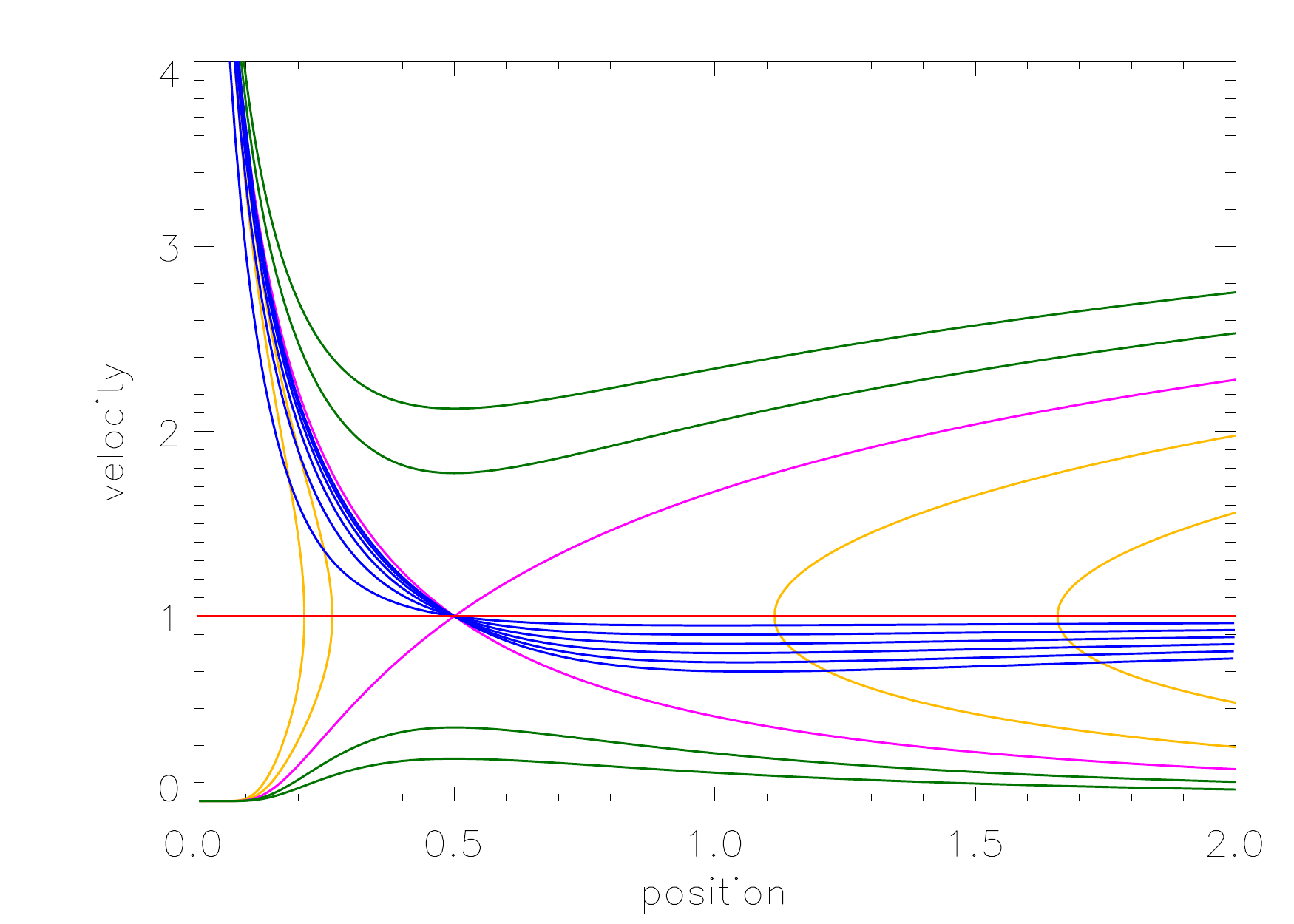}
&
\includegraphics[width=3in]{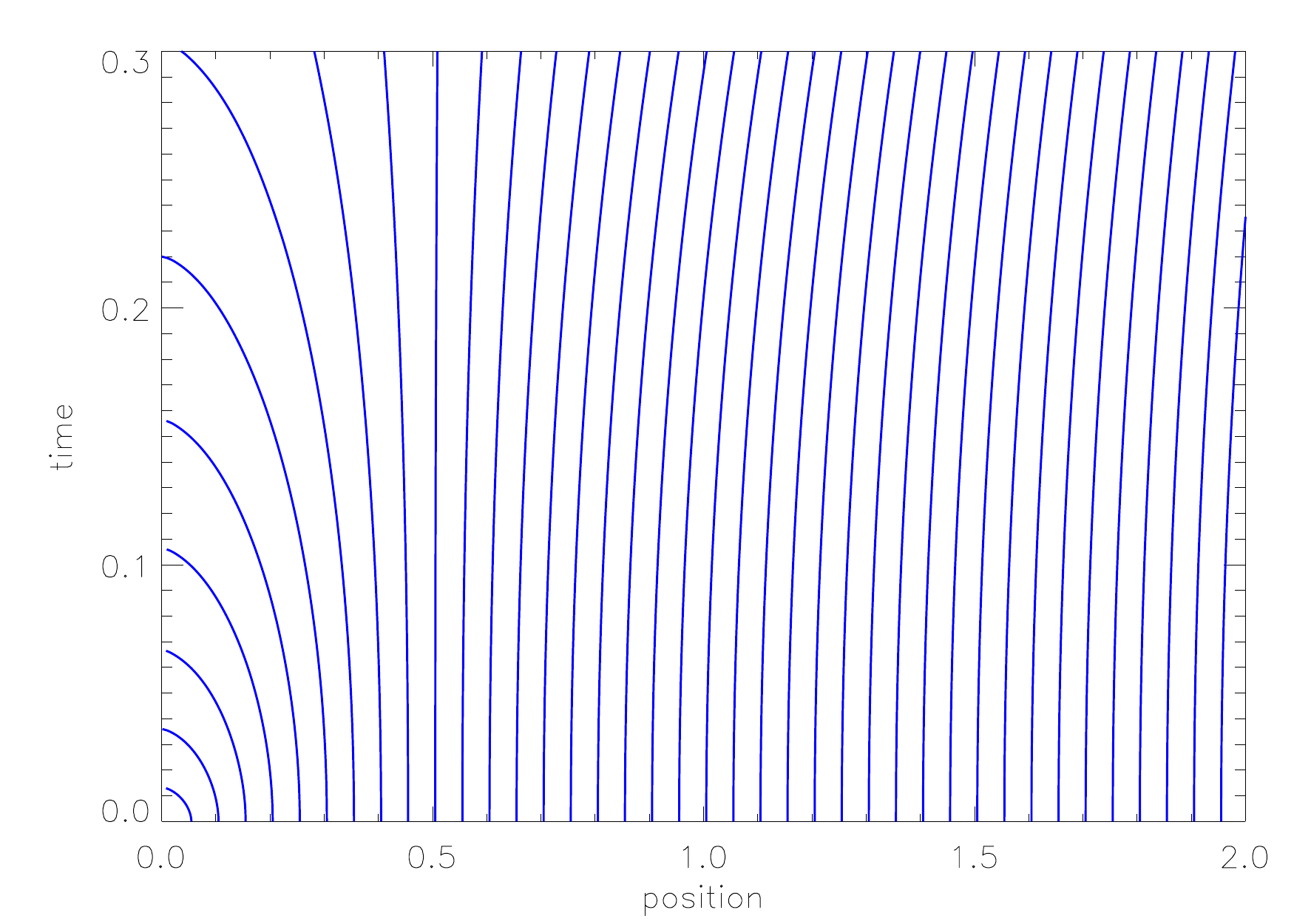}
\end{tabular}
\caption{
Time evolution of an accretion flow with initial values $u(x) = -1.0$ and $0.005 <x < 2.0$
in the same format as figure \ref{bondisub_plot}.
}
\label{bondiss1_plot}
\end{figure*}

\begin{figure*}
\begin{tabular} {p{3in}c}
\includegraphics[width=3in]{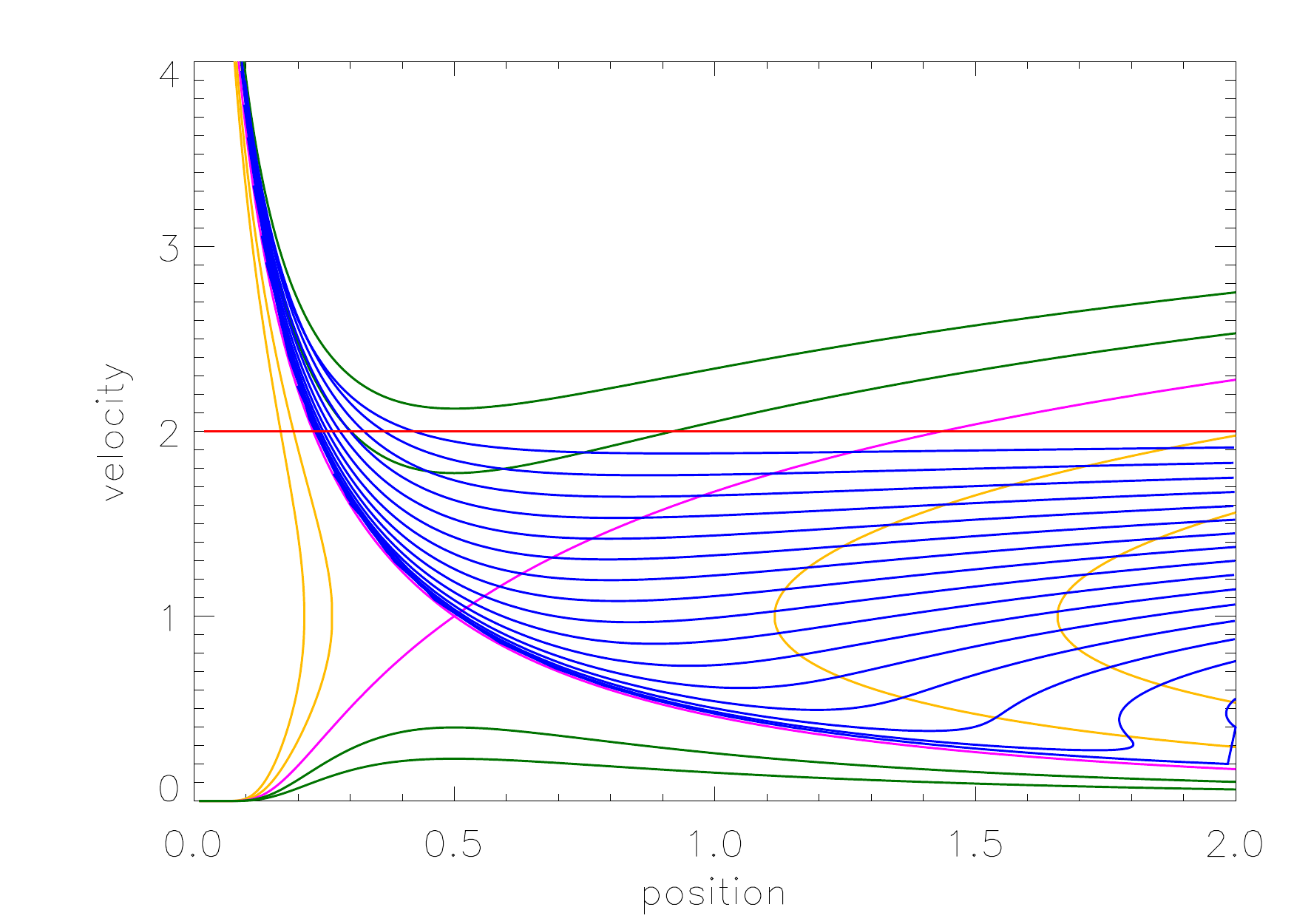}
&
\includegraphics[width=3in]{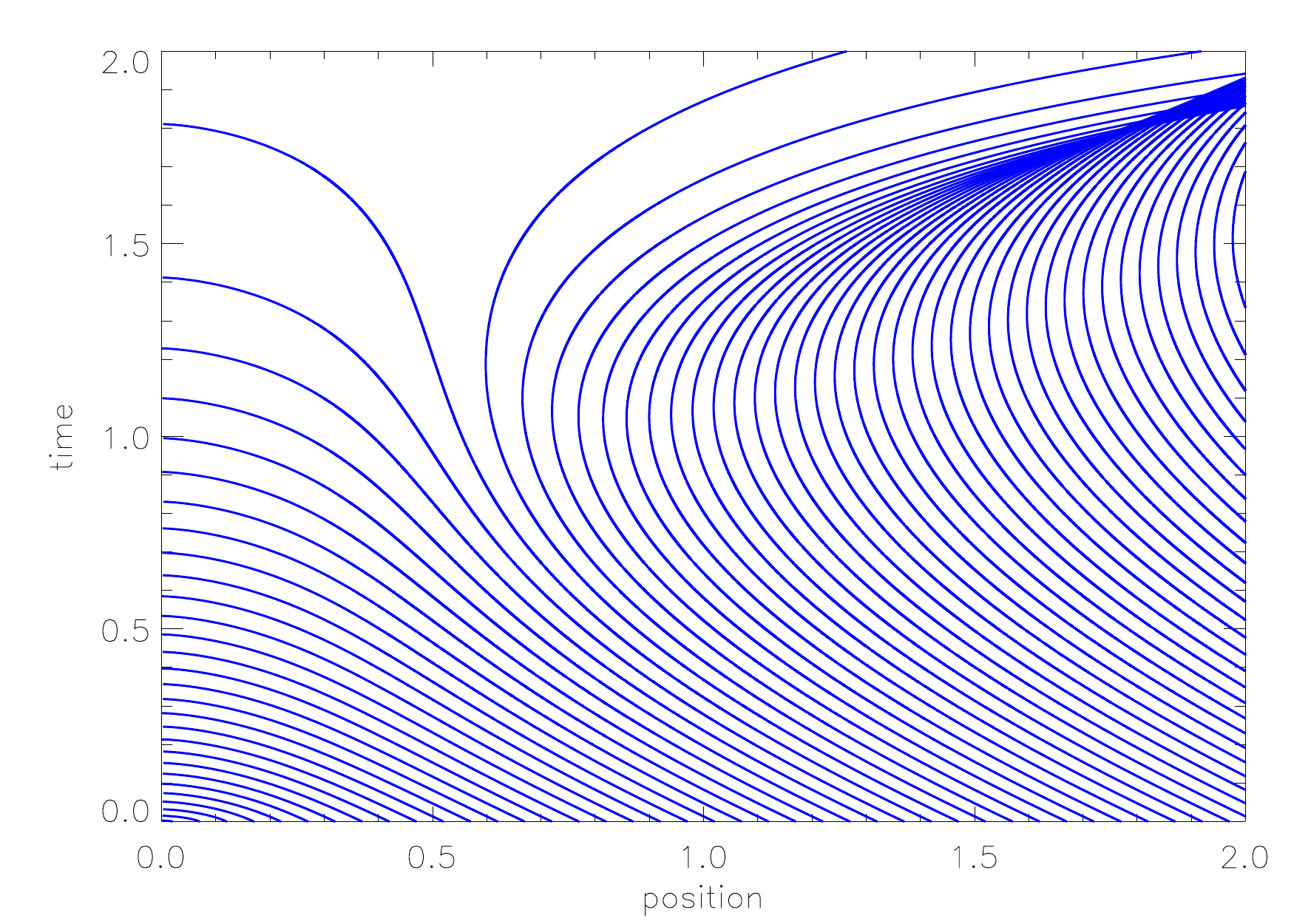}
\end{tabular}
\caption{
Time evolution of an accretion flow with initial values
$u(x) = -2.0$ and $0.005 <x < 2.0$ in the same format as figure \ref{bondisub_plot}.
}
\label{bondiss2_plot}
\end{figure*}

Figures \ref{bondisub_plot} through \ref{bondiss2_plot} show 
the development of a transonic accretion flow starting from
constant initial velocities equal to -0.2, -1.0, and -2.0, respectively, for $0.005 <x < 2.0$. 
(Solutions of equations \ref{lc2} and \ref{lc3} are not defined for initial values of $x=0$ or $u=0$.) 
This evolution is similar to what might be found
in a test of a numerical hydrodynamic simulation evolving to the steady-state starting 
from a constant velocity. In figure \ref{bondiss2_plot}
a shock develops, indicated by multi-valued velocities,
as the velocities outside the critical point decrease to subsonic and are 
impacted by the supersonic initial velocities at larger radii. 
The post-shock velocities evolve to the transonic trajectory.

In the method of characteristics, the
ODE for the velocity $u(t,x(t))$ is integrated along lines of $x(t)$. These lines or characteristics are also shown in the figures.
The point where the characteristics, $x(t)$, first cross indicates the formation of a shock.

\begin{figure*}
\begin{tabular} {p{3in}c}
\includegraphics[width=3in]{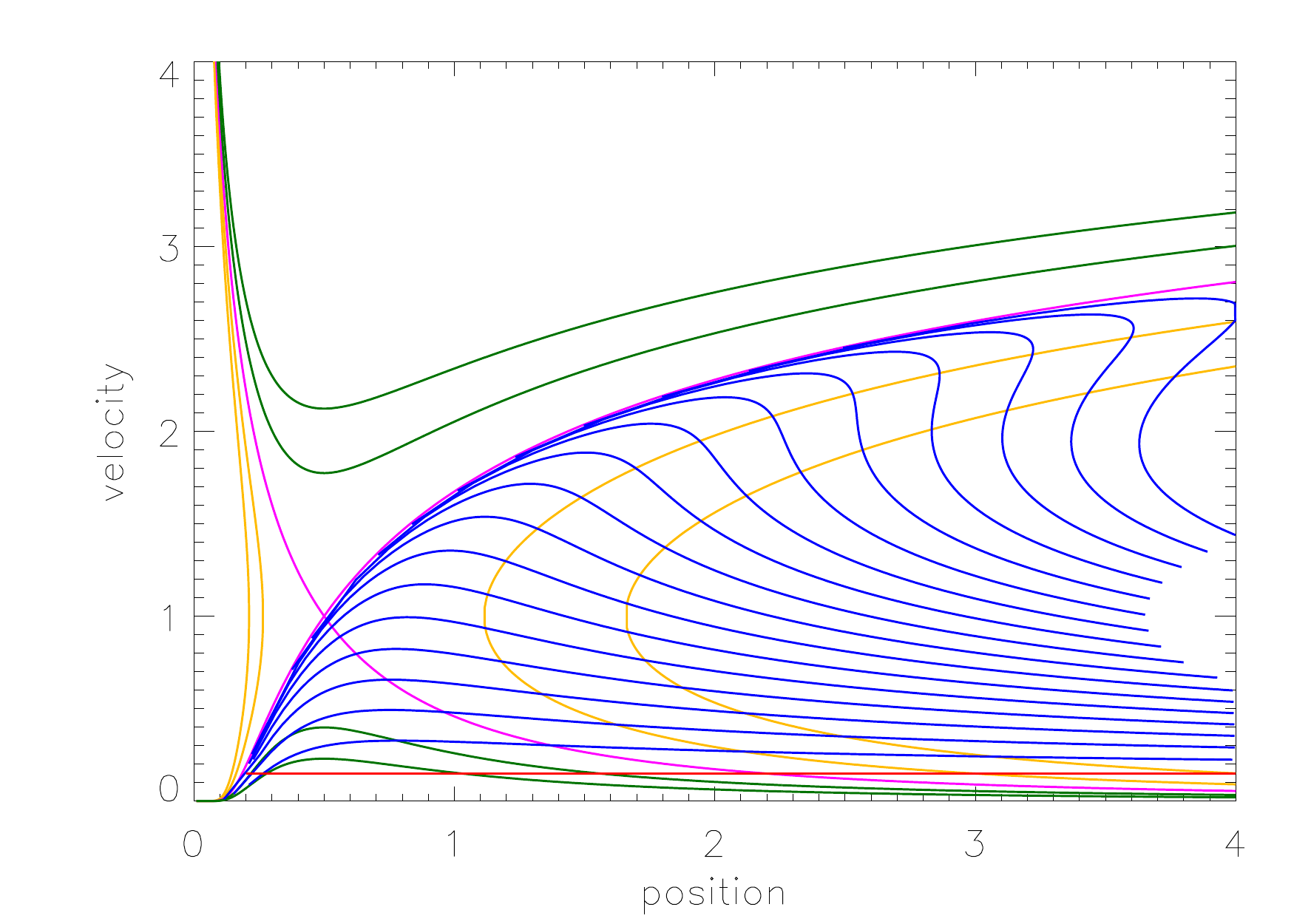}
&
\includegraphics[width=3in]{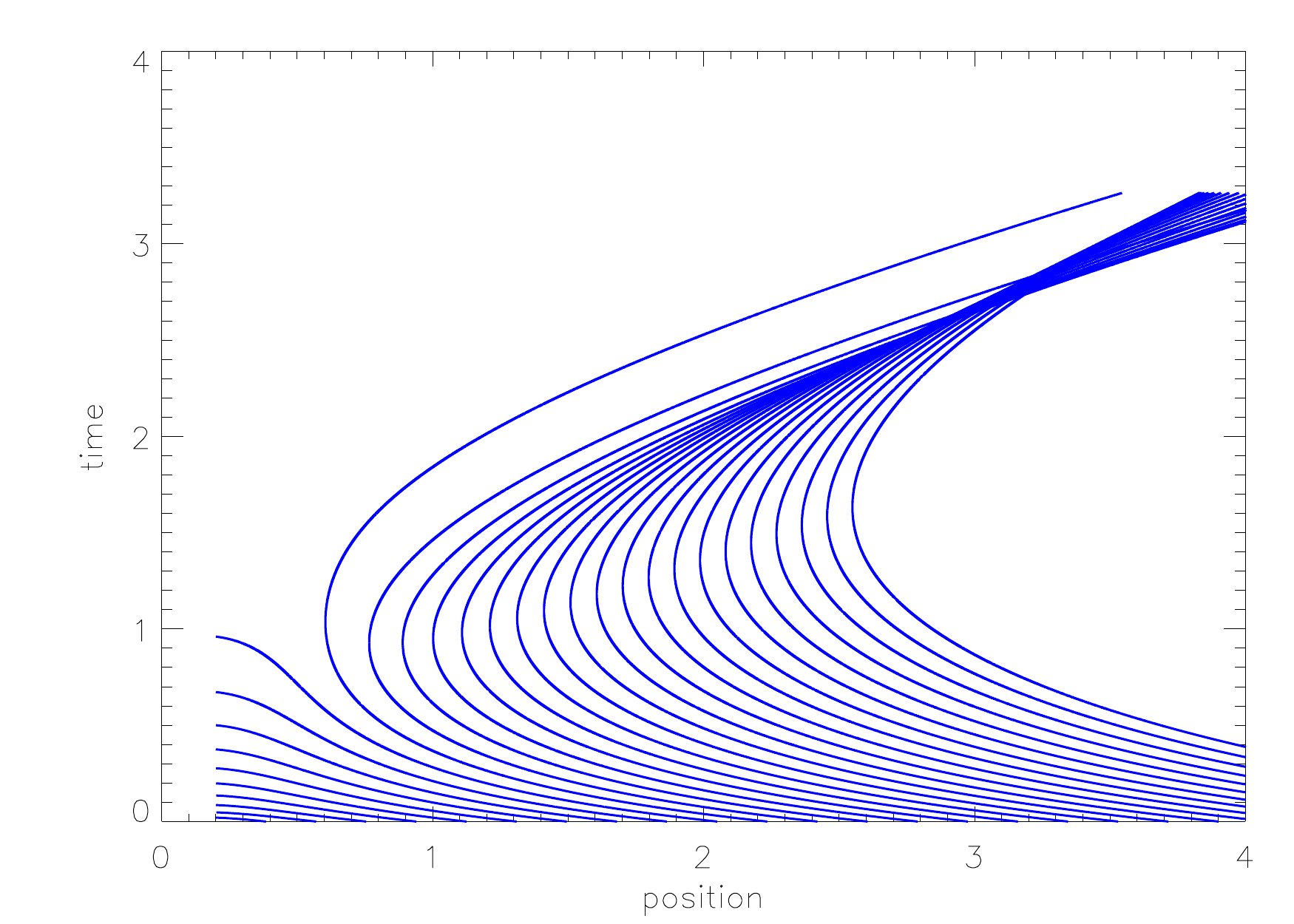}
\end{tabular}
\caption{
Left: Time evolution of a wind with initial values $u(x) = +0.15$ and $0.2 <x < 2.0$ plotted in the same
format as figure \ref{bondisub_plot}. 
}
\label{parkersub_plot}
\end{figure*}

Starting from the same but positive initial velocities, the flows evolve to the outward steady-state 
transonic trajectory,  the Parker wind. 
Figure \ref{parkersub_plot} shows the time evolution for  initial values $u(x) = +0.15$ and $0.2 <x < 2.0$.
A shock develops in the outer region as the outward velocities become supersonic and impact the subsonic
initial velocities 
further out.  Plots
of the evolution from initial velocities, $u(x)=1.0$ and $2.0$ look as expected from figure \ref{parkersub_plot} and are not shown.

The characteristics for the transonic solutions 
in figures \ref{bondisub_plot} 
through \ref{parkersub_plot} have both inward and outward motion.  The location of the point
on the x-axis, $t=0$, that separates
the inward and outward traveling characteristics is the
point where the initial values cross the boundary of a forbidden region. For the case shown in
figure \ref{bondiss1_plot} with initial velocities equal to the sound speed, the point of separation
is the critical point.

\subsection{Subsonic and supersonic solutions}

To find initial values that are everywhere outside the forbidden regions, we start with the
subsonic and supersonic steady-state trajectories.

\begin{figure*}
\begin{tabular} {p{3in}c}
\includegraphics[width=3in]{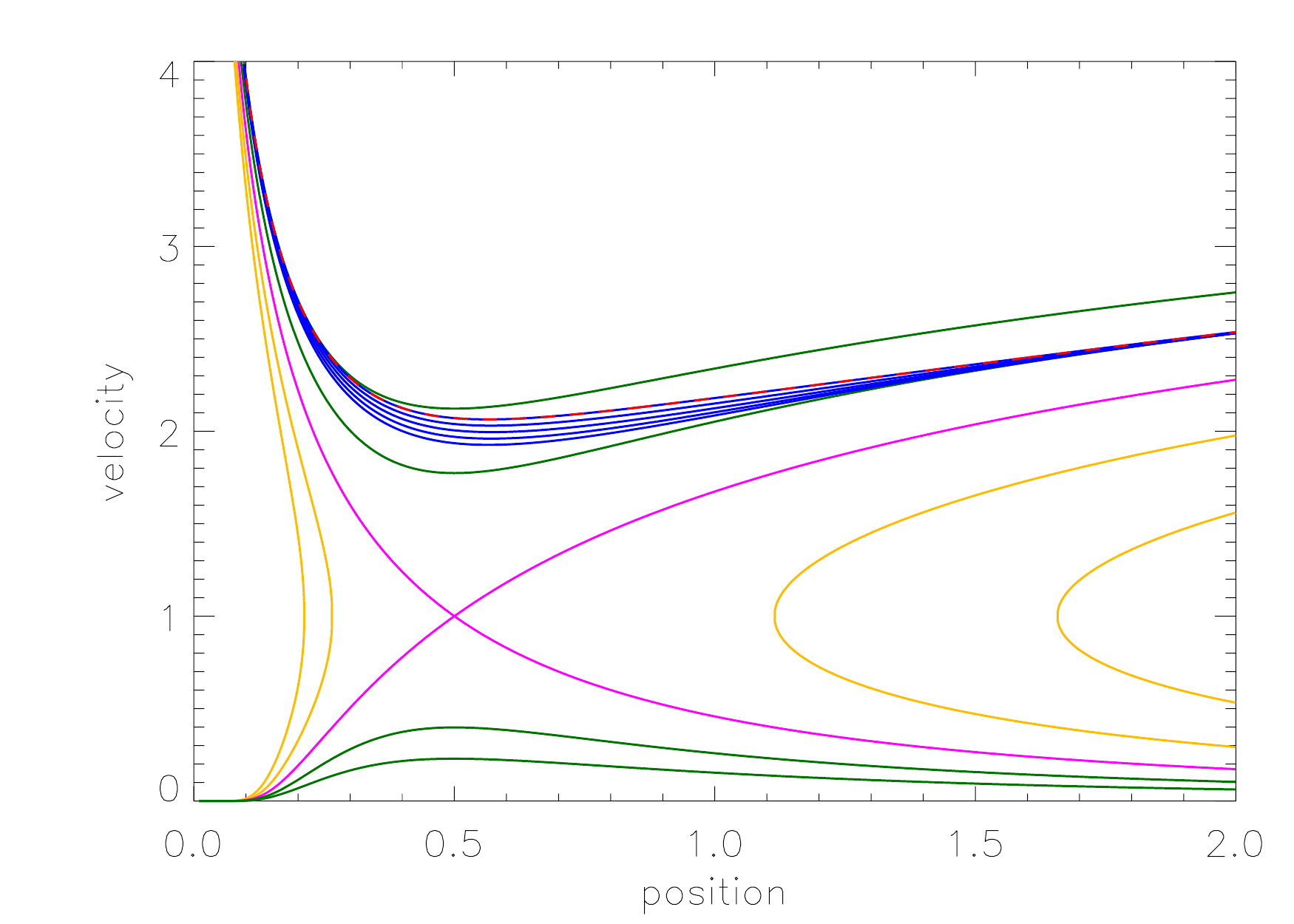}
&
\includegraphics[width=3in]{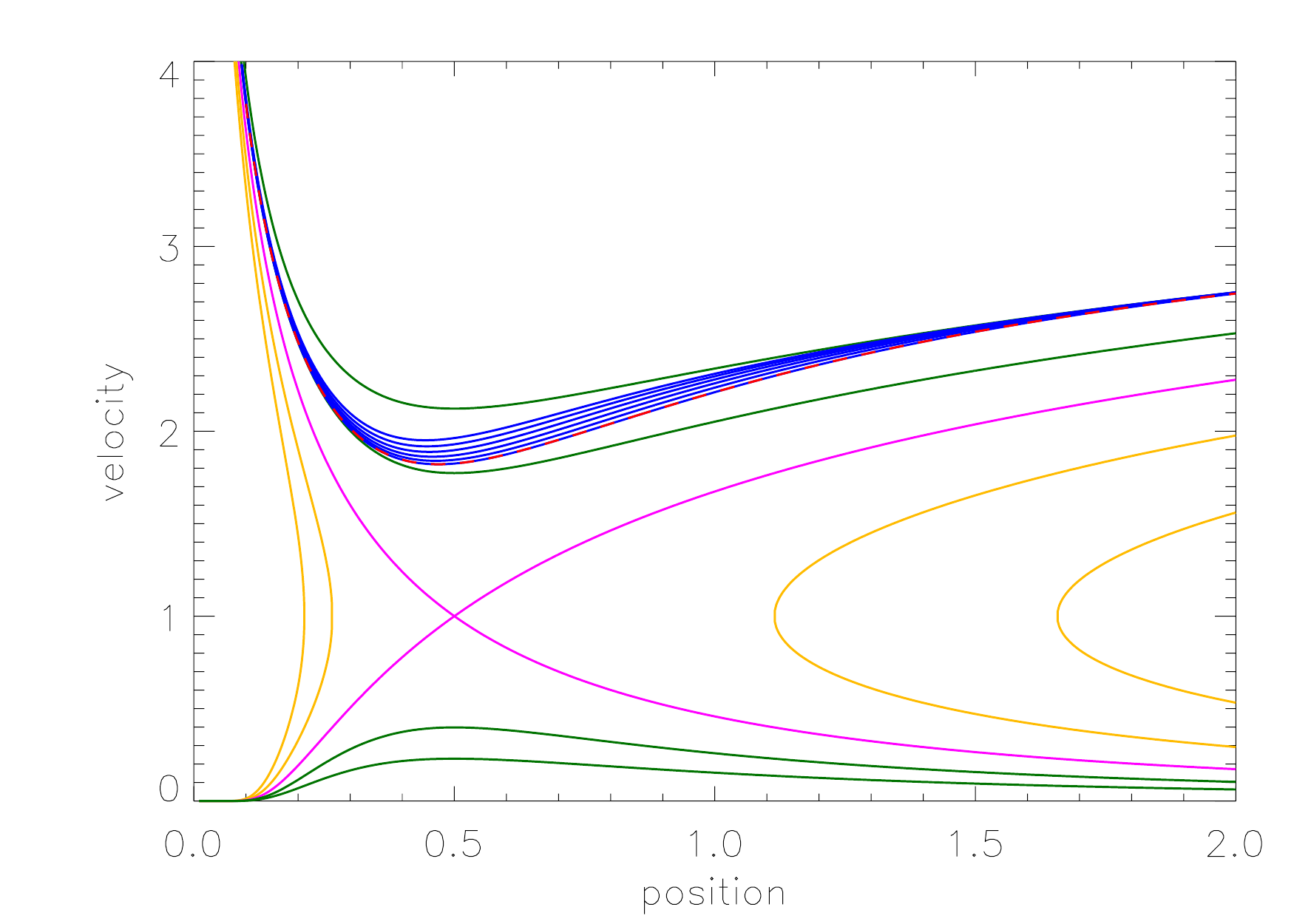}
\end{tabular}
\caption{
Time evolution of a supersonic accretion flow in the same format as figure \ref{bondisub_plot}. The initial values 
are derived as a transition between
two steady-state solutions as explained in the text. The trajectories evolve asymptotically to the steady-state
solution in the outer region.
}
\label{buphilo_plot}
\end{figure*}

Figure \ref{buphilo_plot} shows two examples of 
supersonic accretion. The first begins with initial values that follow the steady-state trajectory with $\mathscr{L}=\mathscr{L}_C+2$
inside the critical point and transition to the steady-state trajectory with $\mathscr{L}=\mathscr{L}_C+1$ outside the critical point by means of a Gaussian.
The second example begins with initial values that transition in the reverse sense. In both cases, these time-dependent flows evolve 
asymptotically to follow the
steady-state trajectory used to set the initial velocities in the outer part of the accretion flow.
While difficult to imagine as an astrophysical flow, the equations may also be solved for the same initial conditions but
changing the initial direction of the flow to a wind.  These solutions evolve to follow the initial steady-state trajectory 
in the inner part of their initial winds. The evolution of these winds is as expected from figure \ref{buphilo_plot} and is not shown.

\begin{figure*}
\begin{tabular} {p{3in}c}
\includegraphics[width=3in]{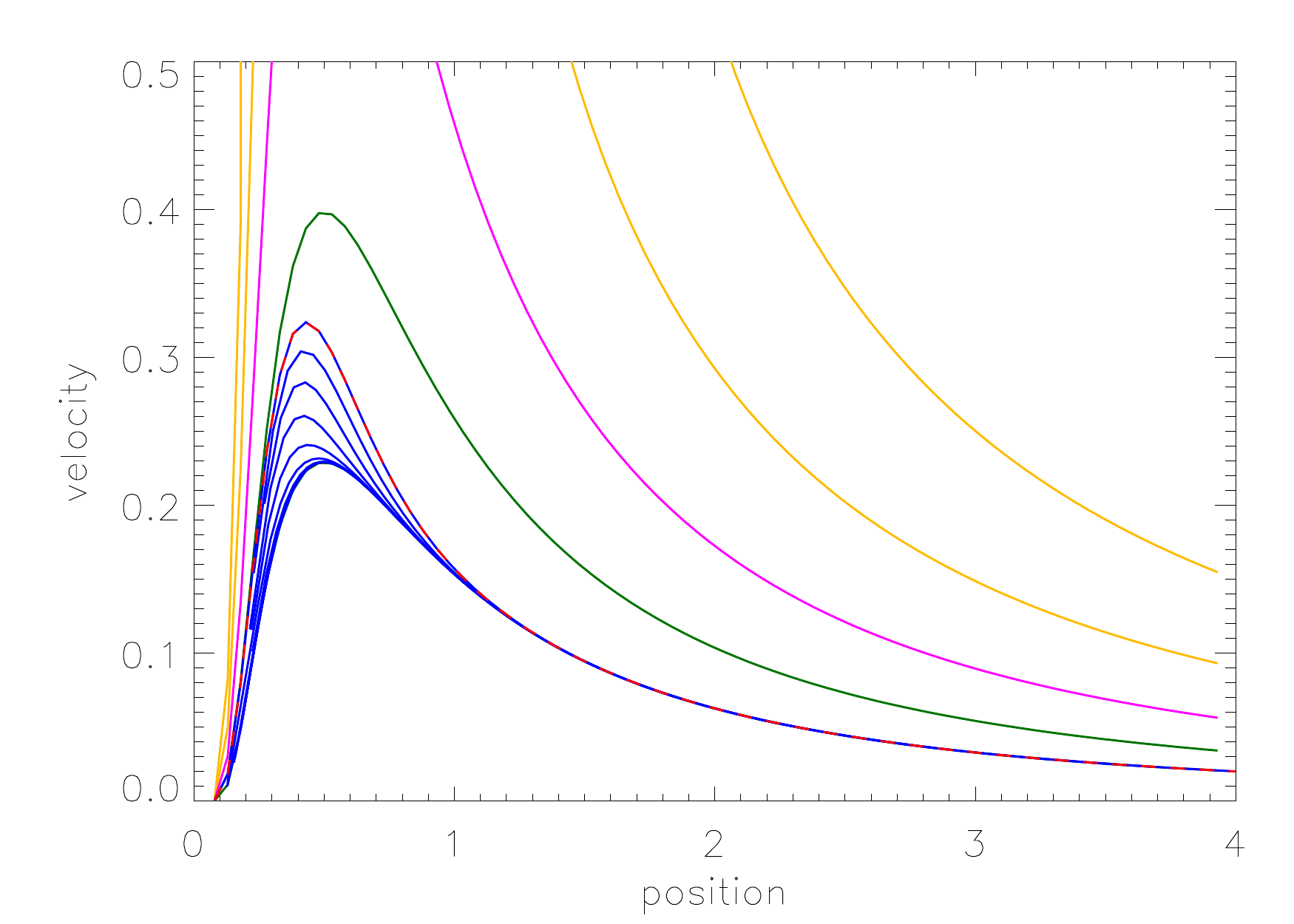}
&
\includegraphics[width=3in]{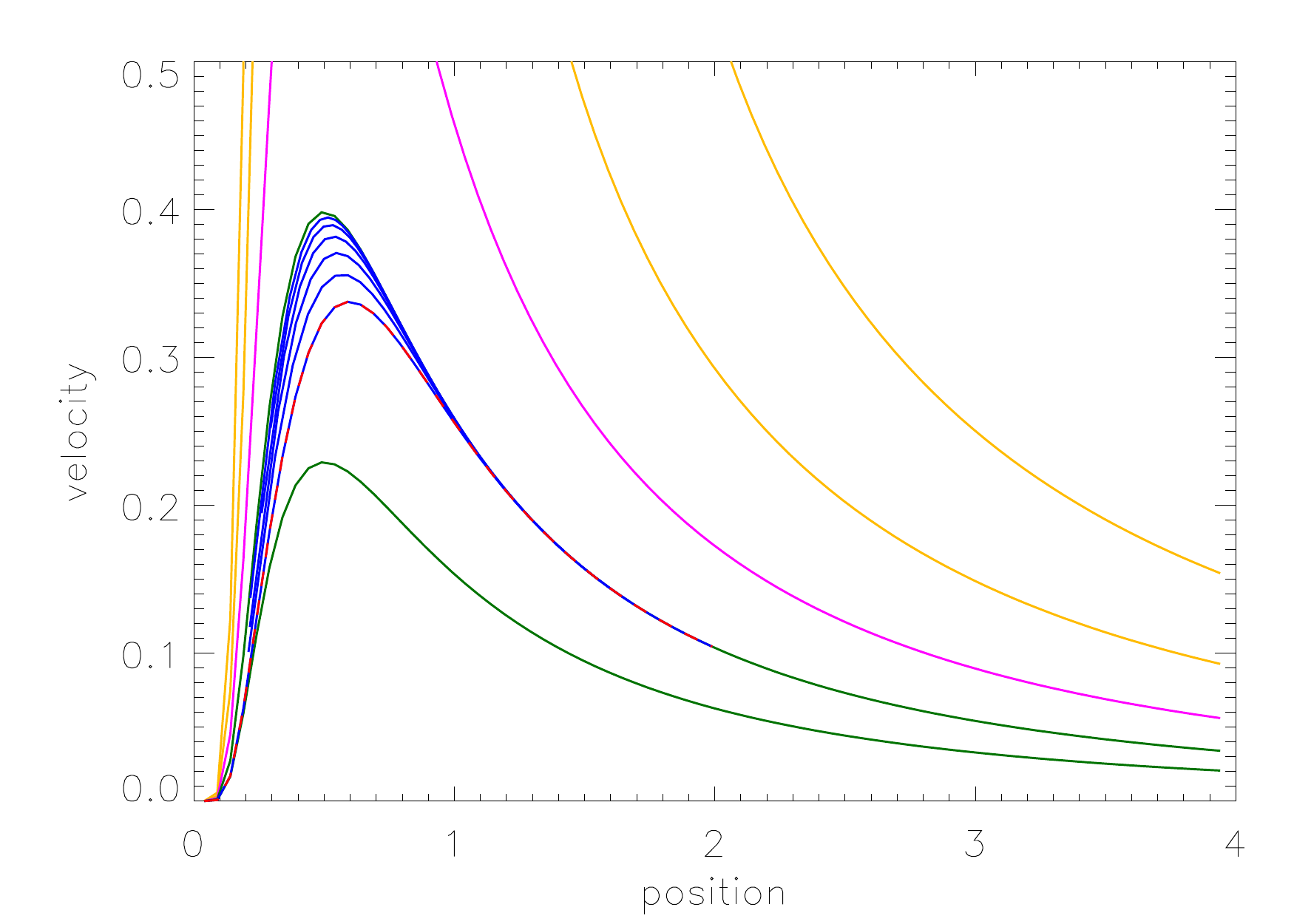}
\end{tabular}
\caption{
Time evolution of a subsonic wind (Parker breeze) in the same format as figure \ref{bondisub_plot}. The initial values 
are derived as a transition between
two steady-state solutions as explained in the text.  The trajectories evolve asymptotically to the steady-state
solution in the outer region.
}
\label{plohilo_plot}
\end{figure*}

\begin{figure*}
\begin{tabular} {p{3in}c}
\includegraphics[width=3in]{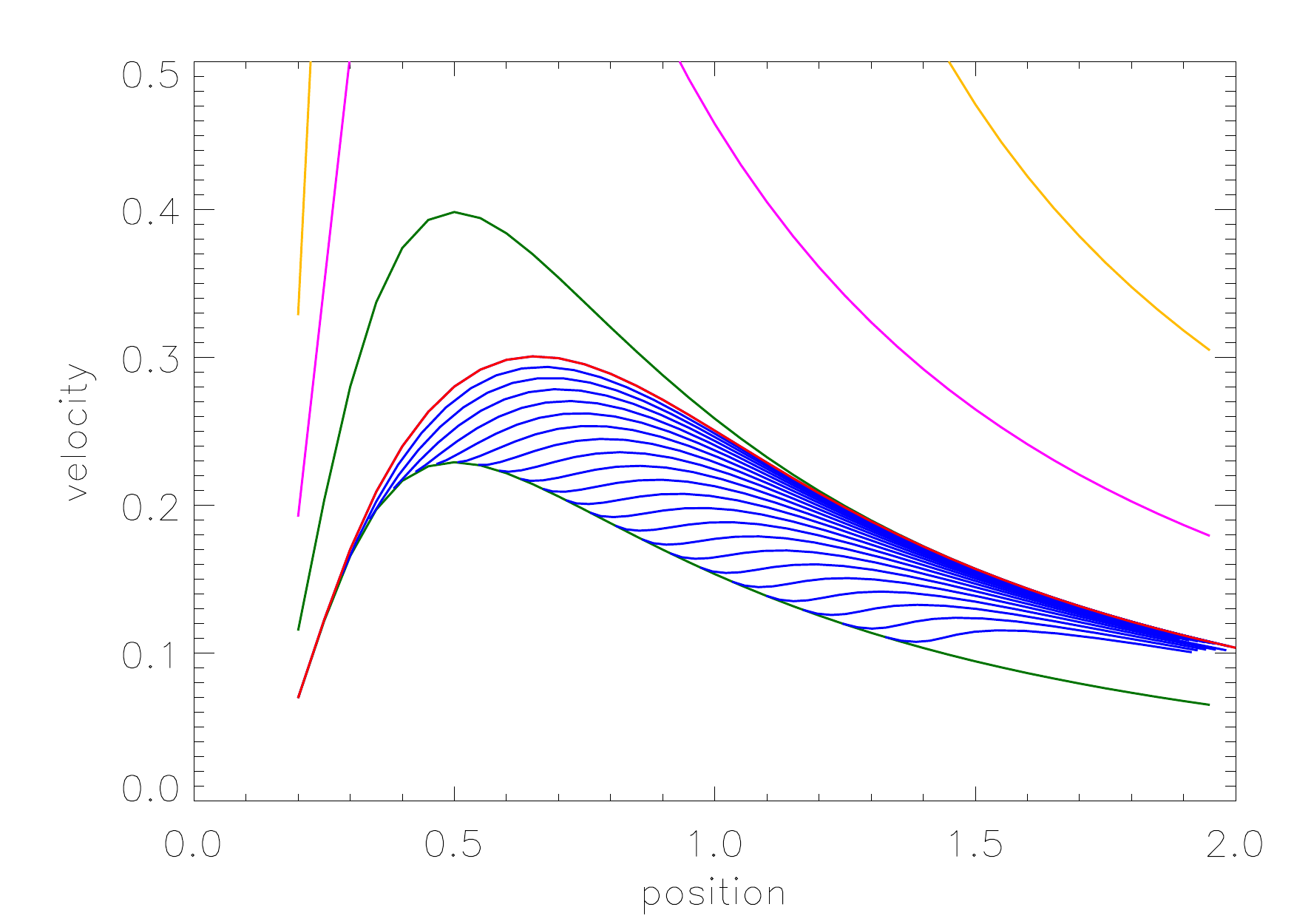}
&
\includegraphics[width=3in]{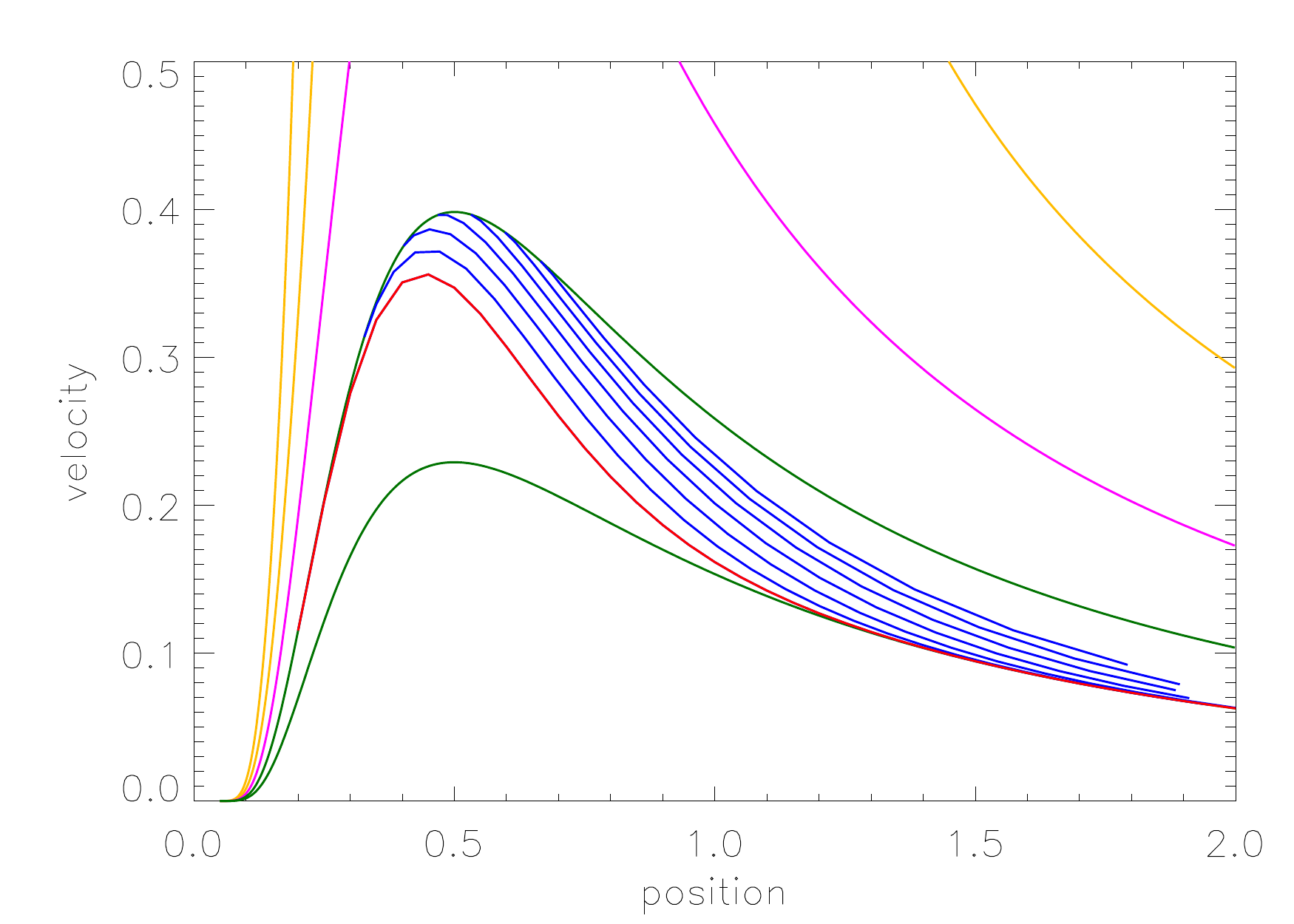}
\end{tabular}
\caption{
Time evolution of a subsonic accretion flow in the same format as figure \ref{bondisub_plot}. The initial values 
are derived as a transition between
two steady-state solutions as explained in the text. The trajectories evolve asymptotically to the steady-state
solution in the inner region. 
}
\label{blolohi_plot}
\end{figure*}

The same calculations can be done for subsonic accretion and winds.
Figure \ref{plohilo_plot} shows a subsonic wind sometimes called a Parker breeze. Both
time-dependent wind solutions evolve asymptotically to follow the initial steady-state trajectories in the outer part of the wind. 
In the case of subsonic accretion, figure \ref{blolohi_plot},
the flows evolve to the initial steady-state trajectories in the inner region rather than the outer
region. 

\begin{figure*}
\begin{tabular} {p{3in}c}
\includegraphics[width=3in]{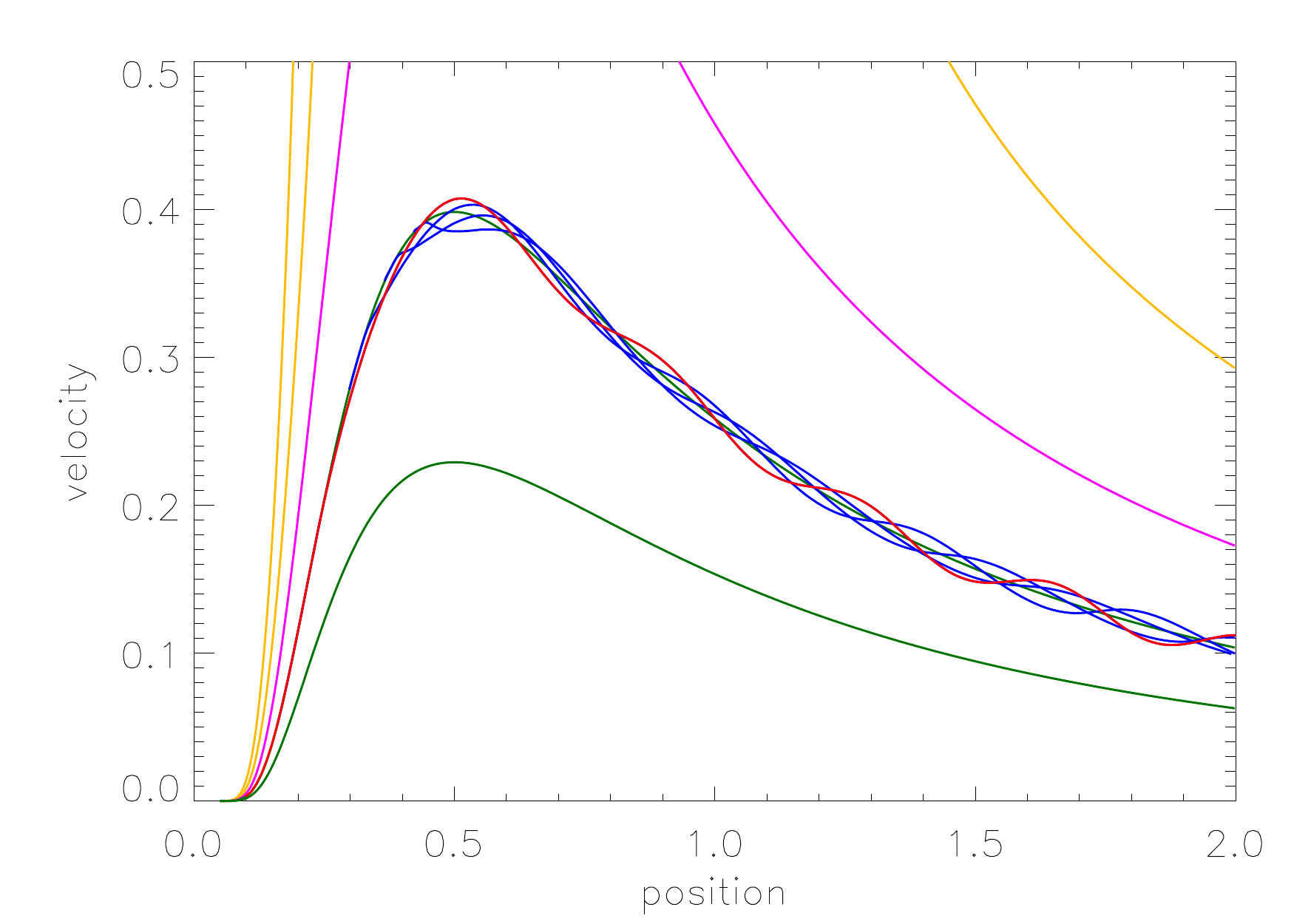}
&
\includegraphics[width=3in]{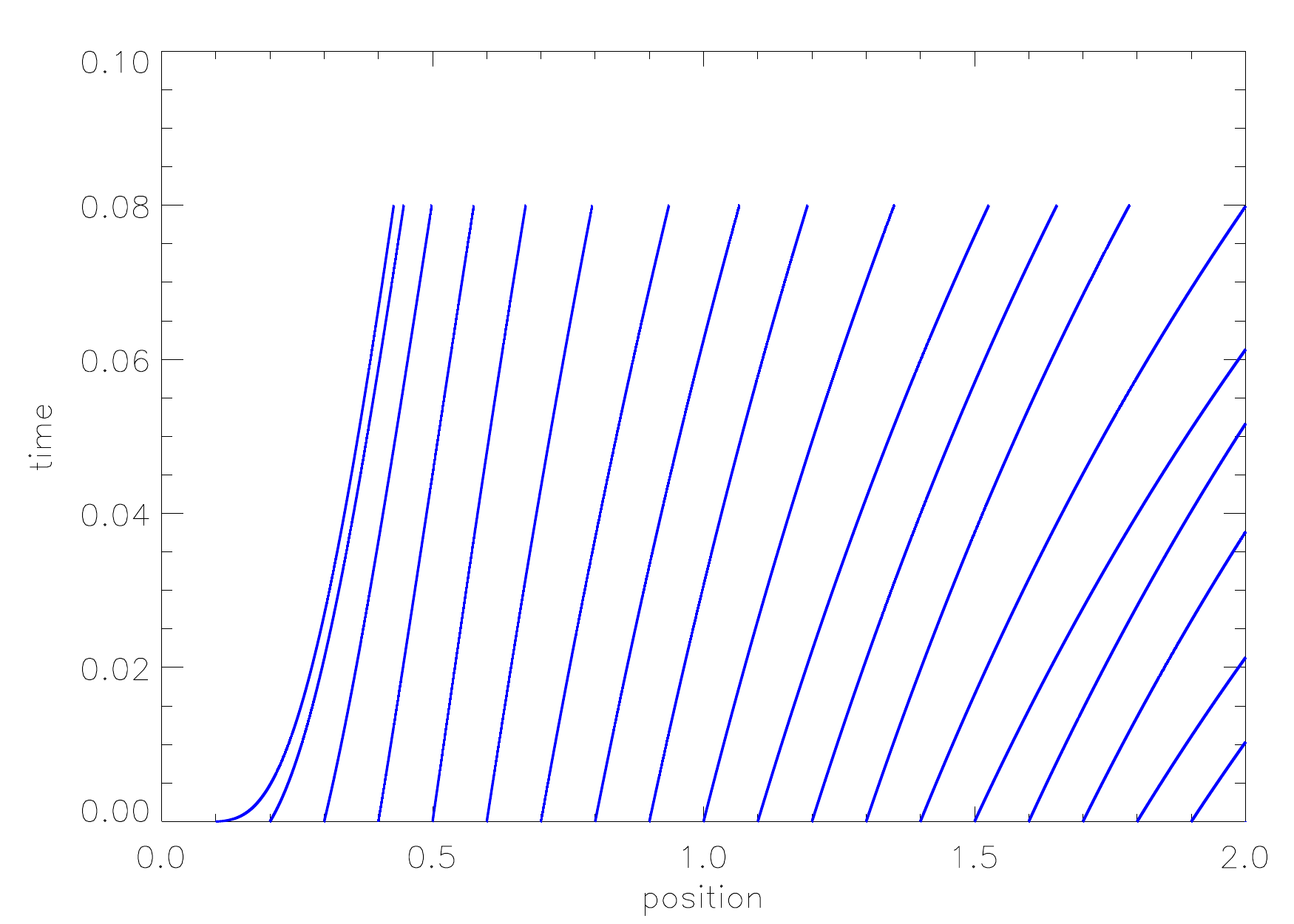}
\end{tabular}
\caption{
Left: Time evolution of a subsonic accretion flow  in the same format as figure \ref{bondisub_plot}. 
The initial values follow a steady-state solution with a sinusoidal perturbation as described in the text.
Right: characteristics for the solution. These indicate that the flow will not evolve out of the subsonic region.
}
\label{bloequi3_plot}
\end{figure*}

As a final example, 
figure \ref{bloequi3_plot} shows the evolution of a subsonic accretion
flow with a sinusoidal perturbation of amplitude 0.05 and period 0.25 imposed on the steady state
between $x=0.25$ and $x=8.0$. (If a finite amplitude
perturbation is continued too close to the origin then some initial velocities would be within the inner
forbidden region and the flow would be swept into approximate free fall allowing the entire flow to 
become transonic.)
The continued oscillation
of the solution shown in the figure is expected because
the density structure in subsonic accretion flows
is approximately hydrostatic and the PDE has no damping. This flow is Lyapunov
stable in the sense that the velocities stay within some range of the initial flow set by the amplitude
of the initial perturbations.

\section{The Barotropic Equation of State}

Since the 
the Parker wind and subsonic accretion both have 
approximately hydrostatic density profiles inside the transonic critical point, comparison with
the stability of hydrostatic equilibrium suggests that the BP flow should be stable with an adiabatic 
equation of state (EOS), 
$\gamma = 5/3$, but not with a
radiation dominated EOS with $\gamma = 4/3$. 
The understanding developed for the isothermal BP flow allows the study to be repeated for
a flow with a more general barotropic EOS, $P \propto \rho^\gamma$.

With a barotropic EOS, the non-dimensional Euler equation equivalent to equation \ref{ndbpeq} is,
\begin{equation} \label{beuler}
\frac {\partial u}{\partial t} = -u \frac {\partial u} {\partial x} - \rho^{\gamma-1} \frac {1} {\rho}   \frac {\partial \rho}{\partial x} - \frac{1}{x^2}.
\end{equation}
Following \citet{Bondi1952}, the steady-state solution,
\begin{equation}
\frac{1}{2}u^2 + \bigg( \frac{1}{\gamma - 1} \bigg) \rho^{\gamma -1} - \frac{1}{\gamma - 1} - \frac{1}{x} = 0,
\end{equation}\label{bss}
can be written in separable variables
similar to equation \ref{ss}
by
scaling the velocity $u$ by the local sound speed.  Substituting
$v = u/a = u \rho^{-(\gamma -1)/2}$ into equation \ref{bss}
along with the continuity equation,
results in
\begin{equation} 
F(v) - \lambda^{ \frac{-2(\gamma-1)}{\gamma+1}}G(x) = 0
\end{equation}
where
\begin{equation}
F(v) = \frac{1}{2} v^{\frac{4}{\gamma+1}} + \frac{1}{\gamma - 1} v^{ \frac{-2(\gamma -1)}{\gamma + 1}} 
\end{equation}
and
\begin{equation}
G(x) = \frac{1}{\gamma-1} x^{ \frac{4(\gamma-1)}{\gamma+1}} + x^{\frac{3\gamma -5}{\gamma+1}} .
\end{equation}
The two terms of the second variation, equivalent to equations \ref{L1} and \ref{L2}, are then,
\begin{equation}
\frac {\partial ^2 G(x) } {\partial ^2 x} = \bigg( \frac{2\gamma-6}{\gamma+1} \bigg) \bigg( \frac {3\gamma -5}{\gamma+1} \bigg) x^{ \frac{\gamma-7}{\gamma-1}}
+  \bigg( \frac{4}{\gamma+1} \bigg) \bigg( \frac{3\gamma-5}{\gamma+1} \bigg)x^{\frac{2\gamma-6}{\gamma+1}}
\end{equation} 
and
\begin{equation}
\frac{ \partial ^2 F(v)}{\partial \dot{x} ^2} = \bigg( \frac{2}{\gamma+1} \bigg)  \bigg( \frac{3-\gamma}{\gamma+1} \bigg) v ^{ \frac {-2(\gamma-1)}{\gamma+1}}
+ \bigg( \frac{2\gamma-2}{\gamma+1} \bigg) \bigg( \frac{3\gamma+1}{\gamma+1} \bigg) v^{ \frac{-4\gamma}{\gamma+1}} .
\end{equation}
The second variation is definite in sign for $\gamma = 5/3$, owing to the factors of $(3\gamma -5)$, implying that an adiabatic BP flow is stable.
Since the factor containing the parameter, $\lambda$, that determines whether the flow is subsonic, transonic, or supersonic, is also multiplied by this zero,
all the trajectories of the adiabatic BP flow are stable.
For $\gamma=4/3$ and most other values, the second variation is indefinite in sign. This does necessarily imply that these flows are unstable,
only that the stability cannot be determined by this Hamiltonian description.

The method of characteristics can also be used to solve for a time-dependent barotropic BP flow. From the PDE \ref{beuler},
the two ODEs equivalent to equations \ref{ode1} and \ref{ode2} can be written for example as,
\begin{equation}
\frac{dx}{dt} = u - \rho^{\gamma-1}\frac{1}{u}
\end{equation}
and
\begin{equation}
\frac{du}{dt} = \rho^{\gamma -1} \frac{2}{x} - \frac{1}{x^2} .
\end{equation}
where $\rho(x,u) $ is given by the continuity equation \ref{continuity}.
Similar to the isothermal case, these equations may be solved parametrically for $u(t,x(t))$. Examples are
best left to specific applications.

\section{Astrophysical Implications}

The steady-state BP flow is particularly useful
as a context to study local phenomena in accretion or winds that involve gas pressure, shocks, or fronts.
In the transonic flows, the critical point along with the conservation of energy completely specifies the
steady-state solution without the need for initial values or 
boundary conditions. The subsonic and supersonic steady-state flows, lacking this constraint, 
require at least one boundary condition, depending on the method of solution, and
both the inner and outer boundaries are problematic.
At the inner boundary, spherical convergence leads to unrealistically
high densities. At the outer boundary, these flows require an asymptotic approach to zero or infinite velocity.
These are all artificial  conditions.  Since 
any other boundary conditions lead to a transonic trajectory, this is
the most likely astrophysical application.

\section{Conclusions}
The Hamiltonian description of the Bondi-Parker flow provides a simple and definitive method for
determining the evolution of time-dependent flows and the stability of the steady state. The method
of characteristics allows a simple solution for the partial differential equation describing the time-dependent
flows as an initial value problem. These methods provide answers to several questions about the
stability and evolution of the flows that were unexplained in Bondi (1954) and Parker (1958). 
In particular, time-dependent flows evolve to the steady state; the steady-state solution for isothermal
and adiabatic equations of state, including
all subsonic, transonic, and supersonic trajectories is at least linearly stable; and a transonic flow
develops under a wide range of realistic initial conditions.

%FIGURE 2

%FIGURE 3

%FIGURE 4

%FIGURE 5
 
%FIGURE 6

%FIGURE 7

%FIGURE 8

%FIGURE 9

\bibliographystyle{mnras}
\bibliography{bondi} % if your bibtex file is called example.bib

%%%%%%%%%%%%%%%%%%%%%%%%%%%%%%%%%%%%%%%%%%%%%%%%%%

%%%%%%%%%%%%%%%%% APPENDICES %%%%%%%%%%%%%%%%%%%%%

%%%%%%%%%%%%%%%%%%%%%%%%%%%%%%%%%%%%%%%%%%%%%%%%%%

% Don't change these lines
\bsp	% typesetting comment
\label{lastpage}
\end{document}